\documentstyle[prl,aps,multicol]{revtex}
\input{epsf}
\begin{document}
\draft
\title{Efficient Quantum Computing of Complex Dynamics}
\author{Giuliano Benenti$^{(a)}$, Giulio Casati$^{(a,b)}$, 
Simone Montangero$^{(a)}$, and Dima L. Shepelyansky$^{(c)}$}  
\address{$^{(a)}$International Center for the Study of Dynamical 
Systems, Universit\`a degli Studi dell'Insubria and} 
\address{Istituto Nazionale per la Fisica della Materia, 
Unit\`a di Como, Via Valleggio 11, 22100 Como, Italy}   
\address{$^{(b)}$Istituto Nazionale di Fisica Nucleare, 
Sezione di Milano, Via Celoria 16, 20133 Milano, Italy}   
\address{$^{(c)}$Laboratoire de Physique Quantique, UMR 5626 du CNRS,
Universit\'e Paul Sabatier, 31062 Toulouse Cedex 4, France}
\date{\today}
\maketitle

\begin{abstract} 
We propose a quantum algorithm which uses the number of qubits in an 
optimal way and {\it efficiently} simulates a physical model with 
rich and complex dynamics described by the quantum sawtooth map.  
The numerical study of the effect of static imperfections in the 
quantum computer hardware shows that the main elements of the 
phase space structures are accurately reproduced 
up to a time scale which is polynomial in the number of qubits. 
The errors generated by these imperfections are more dangerous 
than the errors of random noise in gate operations. 
\end{abstract} 

\begin{multicols}{2}
\narrowtext

When applied to computation, quantum mechanics opens 
completely new perspectives: a quantum computer,
if constructed, could perform certain computations
faster than classical computers, exploiting quantum
mechanical features such as entanglement or superposition
\cite{steane}.
Shor discovered a quantum algorithm \cite{shor} which 
factorizes large integers exponentially faster than any 
known classical algorithm. It was also shown by Grover 
\cite{grover} that the search of an item in an 
unstructured database can be done with a square root 
speed up over any classical algorithm.  
However, at present only a  
few quantum algorithms are known which simulate physical 
systems with exponential efficiency. Such systems  
include certain many-body problems \cite{lloyd}, 
spin lattices \cite{optics}, and models of 
quantum chaos \cite{schack,krot}. 

In this Letter, we present a quantum algorithm which 
computes the time evolution of the quantum sawtooth map,
{\it exponentially faster} than any classical computation.
This model has rich and complex dynamics and finds 
various applications, e.g. for dynamical localization in billiards 
\cite{fausto,prosen,prange}. 
Our algorithm, based on the Quantum Fourier Transform (QFT)
\cite{qft}, simulates the dynamics of a system with $N$ levels 
in $O((\log_2 N)^2)$ operations per map iteration, while a
classical computer, which performs Fast Fourier
Transforms (FFT), requires $O(N\log_2 N)$ operations.
A further striking advantage of the algorithm is the
optimum utilization of qubits: one needs only $n_q=\log_2 N$
qubits (without any extra work space).
We demonstrate that complex phase space structures
can be simulated with less that 10 qubits, while about
40 qubits would allow one to make computations
inaccessible to present-day supercomputers.
This is particularly important, since experiments with
few qubits are being performed at present \cite{ions,nmr},
in particular the QFT was implemented on a three qubit
nuclear magnetic resonance quantum computer \cite{qftexp}. 
For this reason the investigation of this interesting 
physical system will be accessible to first quantum computers, 
operating with few qubits and for which large-scale computations 
like integer factoring are not possible.

The classical sawtooth map is given by
\begin{equation}
\overline{n}={n}+k(\theta-\pi),
\quad
\overline{\theta}=\theta+T\overline{n},
\label{clmap}
\end{equation}
where $(n,\theta)$ are conjugated action-angle variables
($0\le \theta <2\pi$), and the bars denote the variables
after one map iteration. Introducing the rescaled variable
$p=Tn$, one can see that the classical dynamics depends only on
the single parameter $K=kT$, so that the motion is stable for $-4<K<0$
and completely chaotic for $K<-4$ and $K>0$.
For such a discontinuous map the conditions of the
Kolmogorov-Arnold-Moser (KAM) theorem are not satisfied
and therefore, for any $K\ne 0$, the motion is not bounded
by KAM tori. The map (\ref{clmap}) can be studied on
the cylinder ($p\in (-\infty,+\infty)$), which can also be closed
to form a torus of length $2\pi L$, where $L$ is an integer. 
For any $K>0$, one has normal diffusion: 
$<(\Delta p)^2> \approx D(K) t$, where $t$ is the discrete time
measured in units of map iterations and the average $<\cdots>$
is performed over an ensemble of particles with initial momentum
$p_0$ and random phases $0\leq \theta <2\pi$. It is possible to
distinguish two different dynamical regimes \cite{percival}:
for $K>1$, the diffusion coefficient is well approximated by
the random phase approximation, $D(K)\approx (\pi^2/3) K^2$,
while for $0<K<1$ diffusion is slowed down, $D(K)\approx 3.3 K^{5/2}$,
due to the sticking of trajectories close to broken tori
(cantori). For $-4<K<0$ the motion is stable, the phase space
has a complex structure of elliptic islands down to smaller and
smaller scales, and we observed anomalous diffusion,
$<(\Delta p)^2>\propto t^\alpha$, 
(for example, $\alpha=0.57$ when $K=-0.1$).   

The quantum evolution on one map iteration is described
by a unitary operator $\hat{U}$ acting on the wave function
$\psi$:
\begin{equation}
\overline{\psi}=\hat{U}\psi =
e^{-iT\hat{n}^2/2}
e^{ik(\hat{\theta}-\pi)^2/2}\psi,  
\label{qumap}
\end{equation}
where $\hat{n}=-i\partial/\partial\theta$ (we set $\hbar=1$).
The classical limit corresponds to
$k\to \infty$, $T\to 0$, and 
$K=kT=\hbox{const}$. In this quantum model one can
observe important physical phenomena like dynamical
localization \cite{fausto,prosen}. Indeed, due to 
quantum interference effects, the chaotic diffusion
in momentum is suppressed, in a way similar to Anderson localization
in disordered solids. Also in the vicinity of a broken KAM torus,    
cantori localization takes place, since 
a cantorus starts to act as a perfect barrier to quantum wave
packet evolution, if the flux through it becomes less 
than $\hbar$ 
\cite{geisel,mackay,fausto,prosen,prange}. 

The most efficient way to simulate the quantum dynamics
(\ref{qumap}) on a classical computer is based on
forward/backward FFT between $\theta$ and $n$
representations. This is advantageous because the evolution
operator $\hat{U}$ is the product of two unitary operators,
$\hat{U}_k=\exp(ik(\hat{\theta}-\pi)^2/2)$ (kick) and
$\hat{U}_T=\exp(-iT\hat{n}^2/2)$ (free rotation),
which are diagonal in the $\theta$ and $n$ representation,
respectively. Therefore, for a system with $N$ levels,
the one map iteration (\ref{qumap}) requires two FFT and
two diagonal multiplications and can be performed in
$O(N\log_2(N))$ operations.
The dynamics (\ref{qumap}) can be simulated exponentially faster
on a quantum computer with $n_q=\log_2 N$ qubits by means of the
following quantum algorithm:
(i) the wave function
$|\psi\rangle = \sum_{n=0}^{N-1} a_n |n\rangle$
(given in the $n$ representation) is multiplied by $\hat{U}_T$,
so that $\hat{U}_T|\psi\rangle=\sum_n a_n \exp(-iTn^2/2) |n\rangle$.
This step can be done in $n_q^2$ controlled-phase shift gates, as
explained in \cite{krot};
(ii) one can get the wave function in the $\theta$ representation
via the QFT \cite{qft}, which requires $n_q$ single-qubit (Hadamard)
gates and $n_q(n_q-1)/2$ two-qubit gates (controlled phase-shifts);
(iii) the action of $\hat{U}_k$ is diagonal in the angle 
representation and can be simulated in a way similar to (i)
in $n_q^2$ two-qubit gates (we note that this is possible 
thanks to the particular form of $\hat{U}_k$ for the sawtooth map); 
(iv) we go back to the momentum basis performing backward QFT
in $n_q(n_q+1)/2$ gates.
Therefore the whole algorithm requires $n_g=3 n_q^2 + n_q$ quantum
gates per map iteration. 

We model the quantum computer
hardware as a two-dimensional lattice of qubits (spin halves)
with static fluctuations/imperfections in the individual qubit 
energies and residual short-range inter-qubit couplings. The model  
is described by the following many-body Hamiltonian (see also 
\cite{static}):  
\begin{equation}
H_{\hbox{s}}=\sum_i (\Delta_0+\delta_i)\sigma_i^z +
\sum_{i<j}J_{ij}\sigma_i^x\sigma_j^x,
\label{imperf}
\end{equation}
where the $\sigma_i$ are the Pauli matrices for the qubit $i$,
and $\Delta_0$ is the average level spacing for one qubit. 
The second sum in (\ref{imperf}) runs over nearest-neighbor qubit 
pairs, and $\delta_i$, $J_{ij}$ are randomly and uniformly distributed
in the intervals $[-\delta/2,\delta/2]$ and $[-J,J]$,
respectively.
We study numerically the many-body dynamics of the quantum 
computer (\ref{imperf}) running the quantum algorithm 
described above. 
The algorithm is realized by
a sequence of instantaneous and perfect one-
and two-qubit gates, separated by a time interval $\tau_g$,
during which the Hamiltonian (\ref{imperf}) gives unwanted
phase rotations and qubit couplings. 
A special one-qubit rotation is applied to eliminate the 
average phase accumulation given by $\Delta_0$. 
We consider the case
of short range inter-qubit interaction, in which two-qubit
gates can be performed only between nearest neighbor pairs.
As a result, for a square lattice of $n_q$ qubits,
we make $O(n_q^{1/2})$ swap gates to obtain a generic
two-qubit gate. Therefore a single iteration of the map
(\ref{qumap}) requires $n_g=O(n_q^{5/2})$ gates.

\begin{figure} 
\centerline{\epsfxsize=4.2cm\epsffile{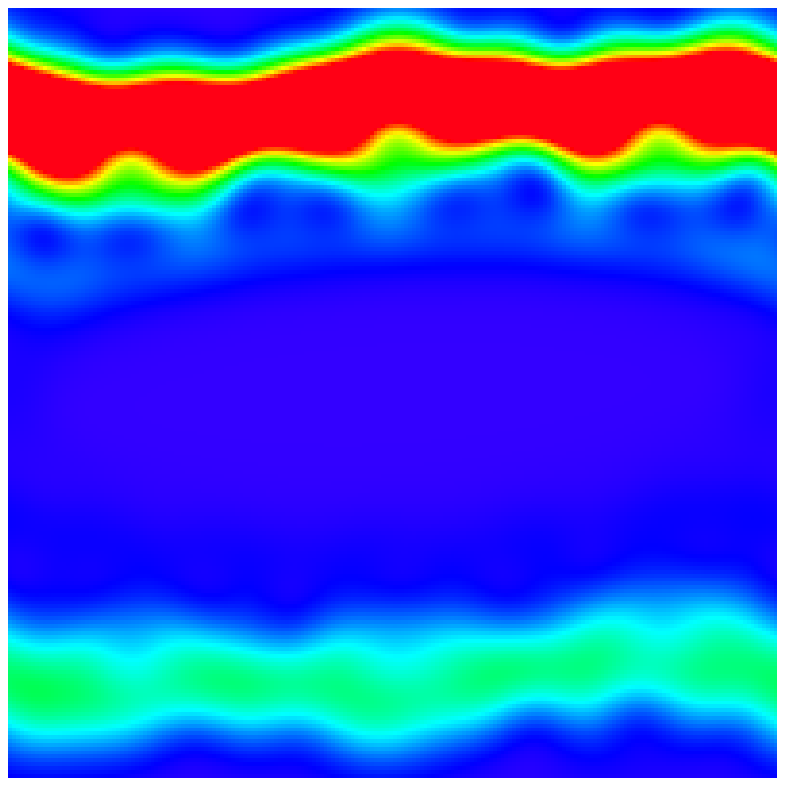}
\hfill\epsfxsize=4.2cm\epsffile{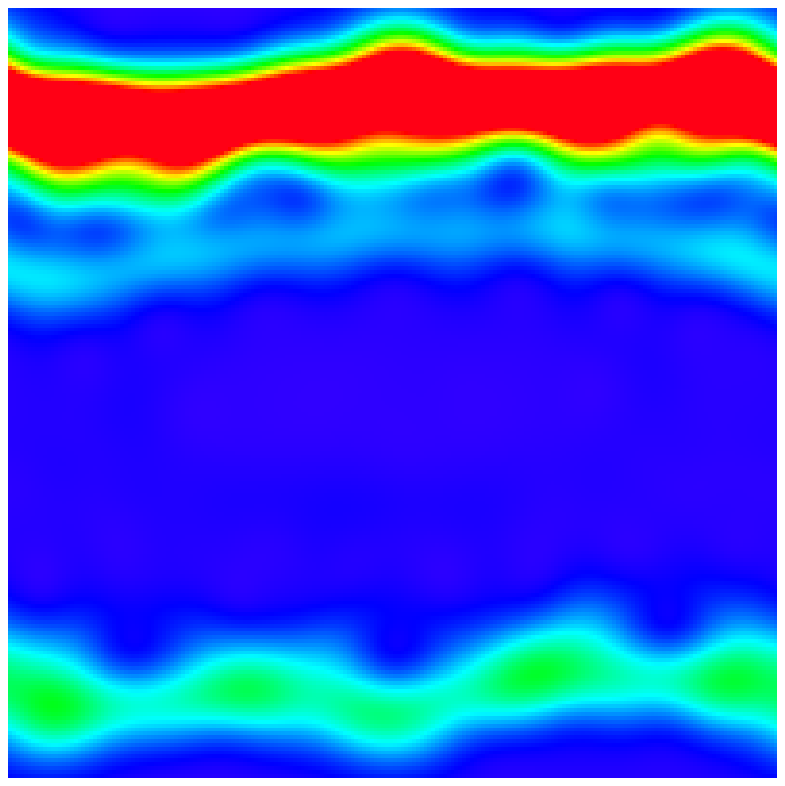}}  
\centerline{\epsfxsize=4.2cm\epsffile{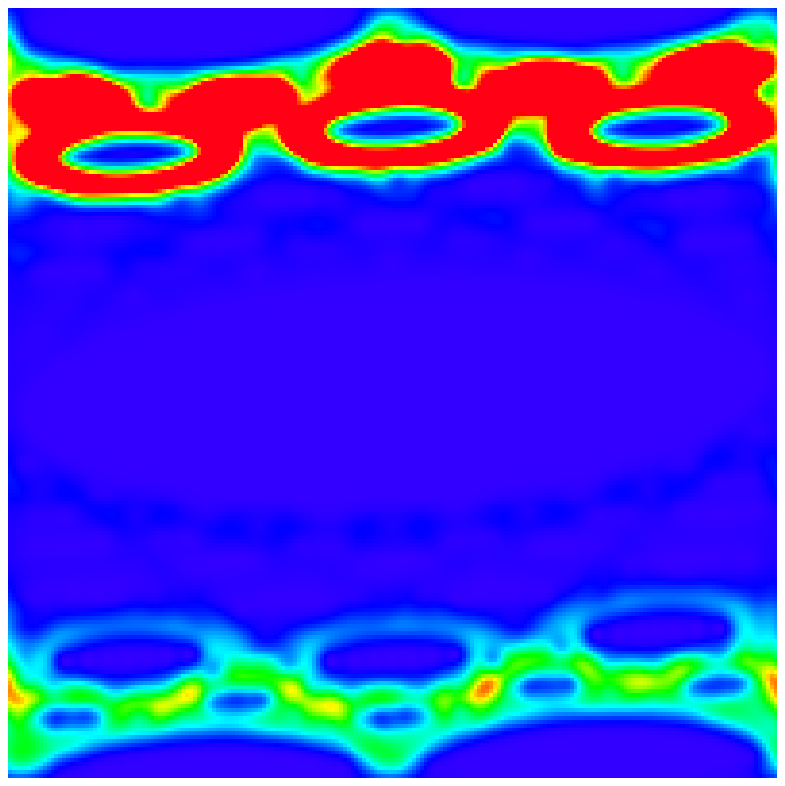}
\hfill\epsfxsize=4.2cm\epsffile{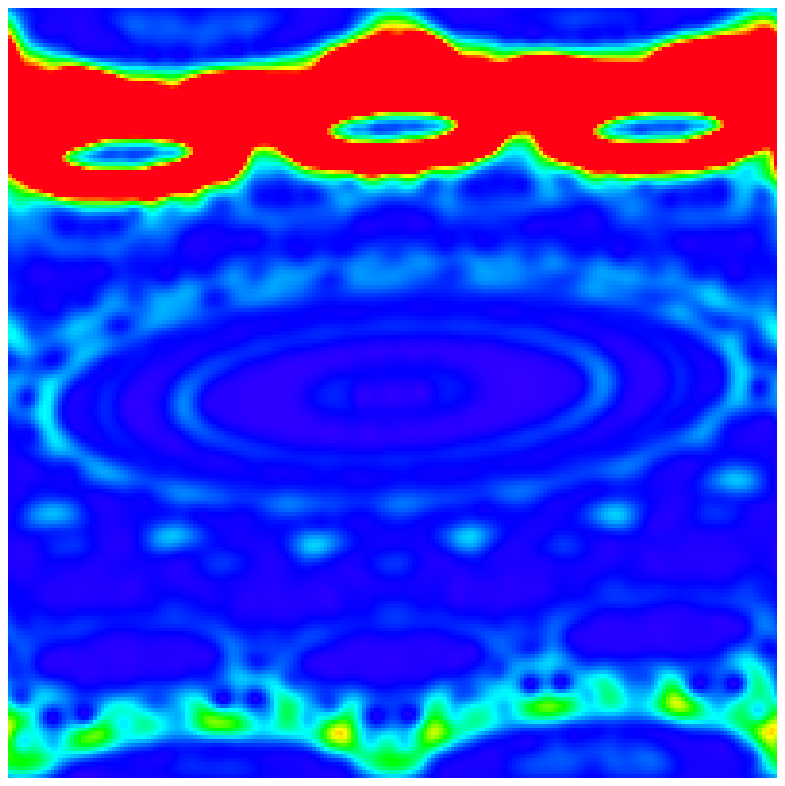}}  
\centerline{\epsfxsize=4.2cm\epsffile{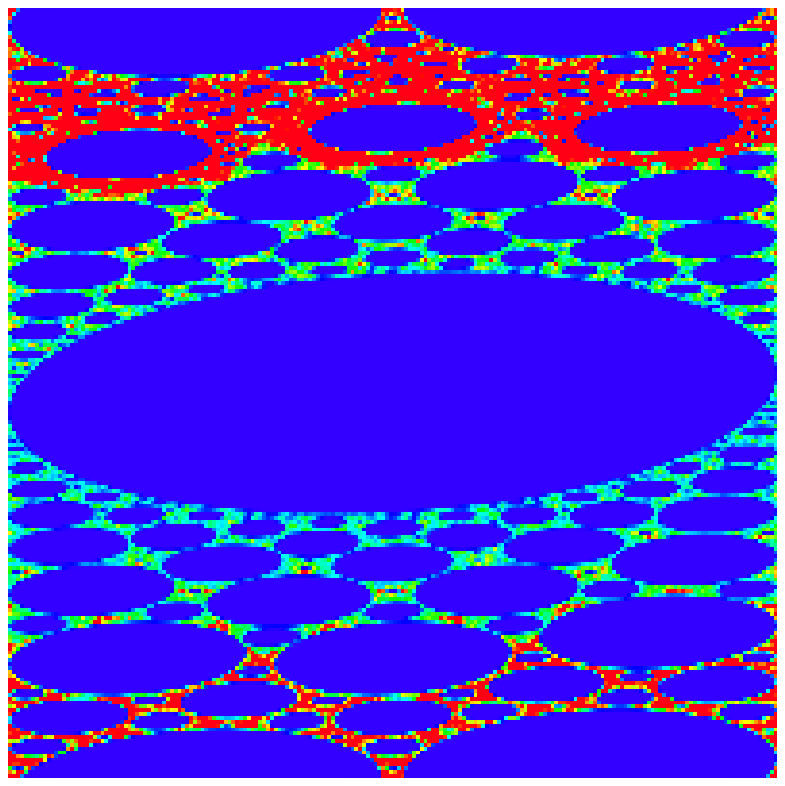}
\hfill\epsfxsize=4.2cm\epsffile{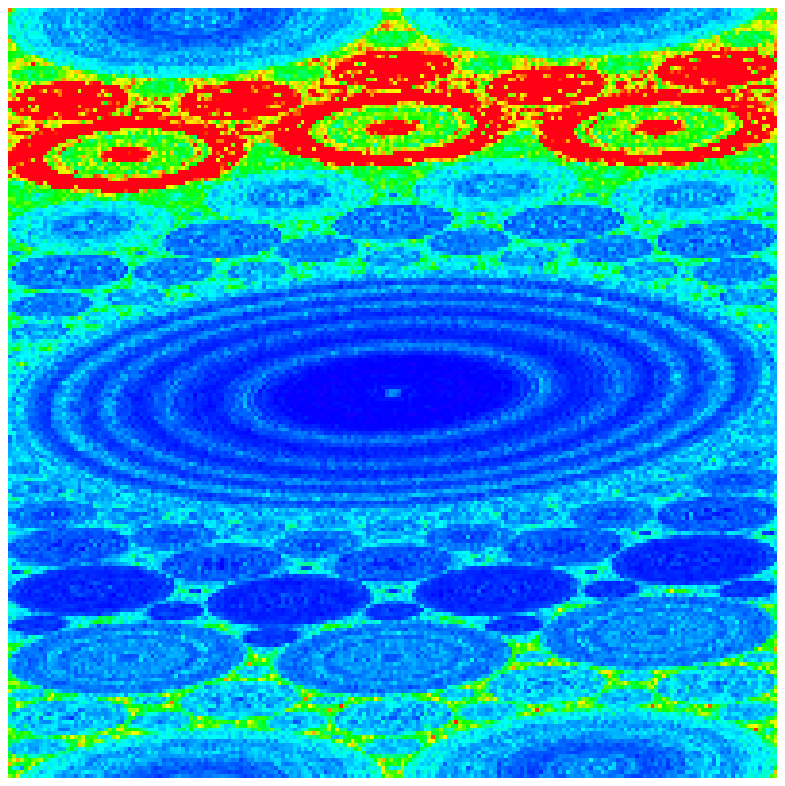}}  
\centerline{\epsfxsize=4.2cm\epsffile{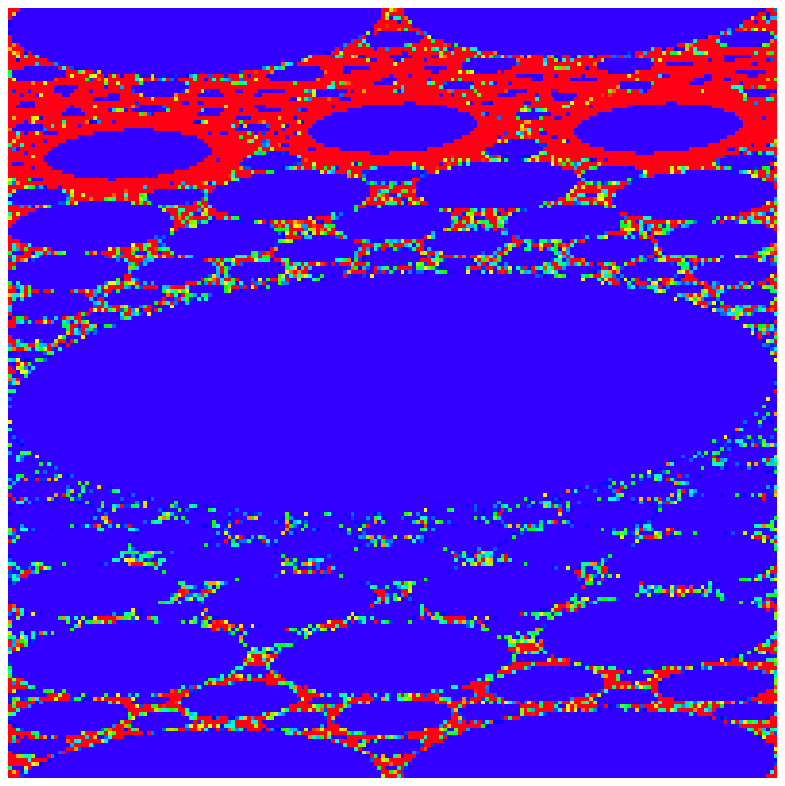}
\hfill\epsfxsize=4.2cm\epsffile{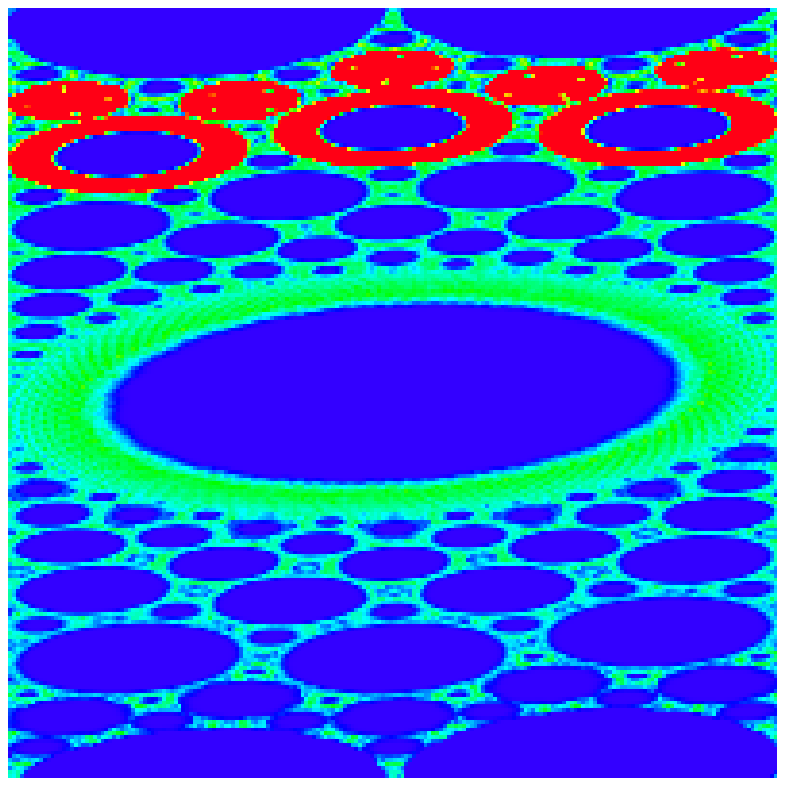}}  
\caption{Husimi function for the sawtooth map in
action angle variables $(p,\theta)$, with 
$-\pi \leq p < \pi$ (vertical axis) and 
$0\leq \theta < 2\pi$ (horizontal axis), for 
$K=-0.1$, $T=2\pi/2^{n_q}$,
$n_0=p_0/T=[0.38 \times 2^{n_q}]$, averaged in the interval 
$950\le t \le 1000$.
From top to bottom: $n_q=6,9,16$ and classical density  
plot, obtained from an ensemble of $10^8$ trajectories,
with initial momentum $p_0=0.38\times 2\pi$ and random 
angles.  
Left and right columns show the case without and with
imperfections: in the quantum case the imperfection 
strength $\epsilon=\delta\tau_g$ scales $\propto n_q^{-3}$,
where $\epsilon=2\times 10^{-3}$ ($n_q=6$), 
$\epsilon=6\times 10^{-4}$ ($n_q=9$), 
$\epsilon= 10^{-4}$ ($n_q=16$), at $J=0$; 
in the classical case round-off errors are of  
amplitude $10^{-3}$.
We choose the ratio of the action-angle uncertainties 
$s=\Delta p/\Delta\theta=1$ ($\Delta p \Delta\theta= T /2$).
The color is proportional to the density: blue for zero and 
red for maximal density. 
}
\label{fig1}       

\noindent 
with $t$ according 
to the law $f(t)\approx \exp(-At^2)$. This can be understood as 
follows: the fluctuations in the 
individual qubit energies, coupled 
to the gate operations in time, give effective Rabi oscillations. 
For this reason $f(t)\sim|\prod_{i=1}^{n_q}\cos(\delta_i\tau_g n_g t)|^2$  
and at short times $f(t)\sim\exp(-n_q (\epsilon n_g t)^2)$.  
This applies also for the case with quantum chaos in the Hamiltonian 
(\ref{imperf}) (e.g. $J=\delta$), which at short time scale can be 
considered from the viewpoint of Rabi oscillations. 
Indeed, here we consider small $\epsilon$ and $n_q$ values, for 
which quantum chaos develops on a chaotic time scale  
$\tau_\chi\approx 1/J\sqrt{n_q}>>\tau_g$ \cite{static}. 
For example, for $n_q=9$, $\epsilon=10^{-4}$, and $J=\delta$ one 
has $\tau_\chi \approx 3\times 10^3 \tau_g$, while one map iteration 
takes time $\tau_1=n_g \tau_g= 413 \tau_g\approx \tau_\chi/10$. 
The fidelity time scale $t_f$ can be obtained 
from the condition $f(t_f)=c$, for example $c=0.9$. The above 
argument gives $t_f\propto \epsilon^{-1} n_g^{-1} n_q^{-1/2} 
\propto \epsilon^{-1} n_q^{-3}$. This estimate is in a good 
agreement with the numerical data shown in Fig.\ref{fig3}. 
These data also show that $t_f$ is not strongly affected 
by the fact that the internal quantum computer hardware
(\ref{imperf}) is in the integrable ($J=0$) or  
quantum chaos ($J=\delta$) regime. 
The most important point is that the dependence of the fidelity time 
$t_f$ on $n_q$ is only polynomial. 

\end{figure}

In this Letter, we study the sawtooth map in the anomalous 
diffusion regime, with $K=-0.1$, $-\pi\leq p <\pi$ (torus geometry).
The classical limit is obtained by increasing the number of qubits
$n_q$, with $T=2\pi/N$ ($k=K/T$, $-N/2\leq n < N/2$). We consider
as initial state at time $t=0$ a momentum eigenstate,
$|\psi(0)\rangle=|n_0\rangle$, with $n_0=[0.38 N]$. Such a state
can be prepared in $O(n_q)$ one-qubit rotations starting from the
ground state $|0,\ldots,0\rangle$. The dynamics of the sawtooth map
reveals the complexity of the phase space structure, as
shown by the Husimi functions of Fig.\ref{fig1}, taken after
$1000$ map iterations. We note that $n_q=6$ qubits are 
sufficient to observe the quantum localization of the anomalous 
diffusive propagation through hierarchical integrable islands.
At $n_q=9$ one can see the appearance of integrable
islands, and at $n_q=16$ the quantum Husimi function explores the 
complex hierarchical structure of the classical phase space. 
The effect of static imperfections (\ref{imperf}) for the operability of the 
quantum computer is shown in Fig.\ref{fig1} (right column).
The data are shown for $J=0$ (we observed similar structures for
$J=\delta$). The main features of the wave packet dynamics 
remain evident even in the presence of significant imperfections, 
characterized by the dimensionless strength $\epsilon=\delta\tau_g$. 
The main manifestation of imperfections is the injection of 
quantum probability inside integrable islands. This creates 
characteristic concentric ellipses, which follow classical 
periodic orbits moving inside integrable islands. These  
structures become more and more pronounced with the increase 
of $n_q$. 
Thus quantum errors strongly affect the quantum tunneling inside 
integrable islands, which in a pure system drops exponentially 
($\propto \exp(-CN)$, $C=$const). 
It is interesting to stress that the effect of quantum errors  
is qualitatively different from the classical round-off errors, 
which produce only slow diffusive spreading inside integrable 
islands (see Fig.\ref{fig1}).  
This difference is related to the fact that spin flips in quantum 
computation can make direct transfer of probability on a large 
distance in phase space. 

\begin{figure} 
\centerline{\epsfxsize=8.cm\epsffile{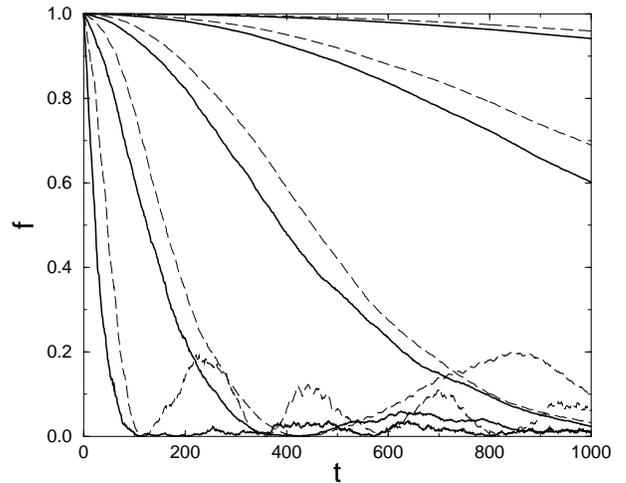}}
\caption{Fidelity as a function of $t$, 
for $n_q=9$ qubits, $J=\delta$ (solid lines) and $J=0$ 
(dashed lines). 
From top to bottom: $\epsilon=10^{-5}, 3\times 10^{-5},
10^{-4}, 3\times 10^{-4}, 10^{-3}$.} 
\label{fig2}
\end{figure}

\begin{figure} 
\centerline{\epsfxsize=8.cm\epsffile{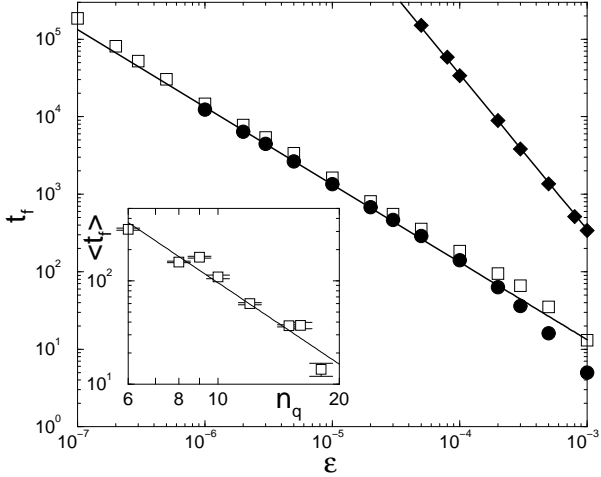}}
\caption{Fidelity time scale $t_f$ 
as a function of $\epsilon$, for $n_q=9$, in the 
case of static imperfections ($J=\delta$ (circles) 
and $J=0$ (squares)) and 
noisy gates (diamonds, see text). 
The straight lines have slopes $-1$ and $-2$.  
The inset shows the dependence of $t_f$ on the number 
of qubits, for $\epsilon=10^{-4}$, $J=0$; the power-law fit 
(straight line) gives $t_f\propto n_q^{-2.6}$. Error bars 
give the statistical errors obtained from 
$10\leq k \leq 500$ random configurations of static 
imperfections.}  
\label{fig3}
\end{figure}

A more quantitative study of the effect of static imperfections 
is obtained from the fidelity of quantum computation, 
defined by $f(t)=|\langle\psi_\epsilon(t)|\psi_0(t)\rangle|^2$,
where $\psi_\epsilon(t)$ is the actual quantum wave function in the 
presence of imperfections and $\psi_0(t)$ is the quantum state for a 
perfect computation. 
The data of Fig.\ref{fig2} show that $f(t)$ drops 
It is interesting to compare the effect of static imperfections 
generated by the Hamiltonian (\ref{imperf}) in the regime of 
perfect gates with the case when the static imperfections are 
absent but the computer operates with noisy gates. To model 
the noisy gates we put $J=0$ and $\delta_i$ fluctuating 
randomly and independently from one gate to another in the 
interval $[-\delta/2,\delta/2]$. In this case the fidelity 
drops in a qualitatively different way $f(t)\approx\exp(-Bt)$ 
(data not shown). This corresponds to the Fermi 
golden rule regime, where the probability at a given state 
decays exponentially with time. At each gate operation a  
probability of order $\epsilon^2$ is transferred to other 
states, so that the fidelity time scale is given by 
$t_f\sim 1/(\epsilon^2 n_g)$. This estimate is in good 
agreement with numerical data shown in Fig.\ref{fig3}. 
We found the same behavior for $J=\delta=0$ and noisy gates with 
unitary rotations on a random angle $\epsilon$. 
The dependence on $\epsilon$ 
is qualitatively different comparing to the static imperfection 
case (\ref{imperf}) discussed above. We stress that static 
imperfections give shorter times $t_f$ and therefore are more 
dangerous for quantum computation.    
We note that the quadratic and linear regimes of fidelity 
decrease with time were respectively discussed in Refs. 
\cite{zurek1,zurek2}.  

In summary, we have shown that a realistic quantum computer can simulate
with exponential efficiency complex dynamics with rich phase
space structures. Our studies demonstrate that the main
structures inside the localization domain are rather stable
in the presence of static imperfections.
We show that the errors from static imperfections are stronger 
than the errors of noisy gates. 
Of course it is not possible to extract all exponentially 
large information hidden in the wave function with $2^{n_q}$ 
states. However, it is possible to have access to coarse 
grained information. For example from a polynomial number of 
measurements one can obtain the probability distribution over 
momentum (or angle) states. 
This allows one to study the anomalous diffusion in the 
deep semiclassical regime. 
Such an information is not accessible 
for classical computers which cannot simulate more than 
$2^{40}$ quantum states.  
Recently an efficient algorithm was proposed in Ref. 
\cite{saraceno}, which allows one to measure the value  
of the Wigner function at a chosen phase space point. 
This can also provide important new informations about 
quantum states in systems with hierarchical phase space 
structures.  
We believe that in the near future our results can be 
experimentally observed in quantum computers 
operating with a small number of qubits. 

This work was supported in part by the EC RTN network contract 
HPRN-CT-2000-0156 and (for D.L.S.) by the NSA and ARDA under 
ARO contract No. DAAD19-01-1-0553. Support from the PA 
INFM ``Quantum transport and classical chaos'' is 
gratefully acknowledged.

\end{multicols}

\end{document}